\documentstyle[preprint,aps]{revtex}

\begin{document}
\title{Photoinduced Superconducting Nanowires in GdBa$_2$Cu$_3$O$_{6.5}$ films.}
\author{R. S. Decca,$^{\dag}$H. D. Drew,$^{\dag}$B. Maiorov,$^{\ddag}$  J. Guimpel,$^{\ddag}$  
and E. Osquiguil$^{\ddag}$  }
\address{$^{\dag}$ 
Laboratory for Physical Sciences
and Department of Physics, University of Maryland, College Park, MD 20742.\\
$^{\ddag}$ Centro At\'{o}mico Bariloche and Instituto Balseiro, \\Comisi\'on Nacional de
Energ\'{\i}a
At\'omica, 8400 S. C. de Bariloche, R. N., Argentina.}
\date{\today}
\maketitle
\begin{abstract}
We report the fabrication of high T$_{c}$~superconducting wires by
photodoping a GdBa$_2$Cu$_3$O$_{6.5}$ thin film. An optical near-field
probe was used to locally excite carriers in the system at room temperature.
Trapping of the photogenerated electrons define a confining potential for
the conducting holes in the CuO planes. Spatially resolved
reflectance measurements show the photogenerated nanowires to be $\sim $ 
250 nm wide. Electron diffusion, before electron capture, is believed to be
responsible for the observed width of the wires.
\end{abstract}


One of the most important prospects with high temperature
superconductors is their application in electronic circuits.\cite{aplica}
Therefore the capability to define narrow lines and small Josephson
junctions is critical for the technological development of these materials.
Many applications would benefit from submicron feature
sizes, not attainable with conventional patterning and etching.\cite{aplica} 
An alternative method to define superconducting wires is suggested by
the persistent photoconductivity (PPC)\cite{PPC,osqui1} and
persistent photoinduced superconductivity (PISC)\cite{gladys} effects observed in
oxygen-deficient {\it R}Ba$_{2}$Cu$_{3}$O$_{6+x}$ (RBCO) superconductors. It
has been shown that under visible\cite{osqui1} and ultraviolet\cite{schuller} 
illumination 
photodoping closely resembles the more common oxygen order self 
doping in RBCO, increasing  
the carrier density.\cite{hall} The relaxation time of
photogenerated carriers is several hours at room
temperature, and becomes persistent if the sample is kept at temperatures
below 100 K.\cite{osqui1} Recent experiments have shown
that electron detrapping involves oxygen ion movement,\cite{julio}
suggesting that captured electrons will be localized on a scale of order of
the dimensions of the unit cell. Since all of the previous experiments were carried out
with broad area illumination, this small localization length has
not been experimentally exploited. 

In this Letter we show that photocarriers can be induced and confined on a
scale of 250 nm by illuminating the sample with 
a near-field scanning optical microscope (NSOM)\cite{NSOM} probe. 
A {\it c}-axis oriented, 180 nm thick GdBa$_{2}$Cu$_{3}$O$_{6+x}$ (GBCO) 
thin film was grown on a (100)
MgO substrate by dc magnetron sputtering, as described elsewhere.\cite
{julio} The as-grown film shows linear temperature dependence for
the DC resistivity a T$_{c}$~of 89.4 K, with a transition width of 1 K, as measured by AC 
susceptibility measurements. The oxygen content of the sample was adjusted to x=0.5, and 
the sample was
patterned by conventional photolithography and wet etching into a four probe
geometry.\cite{julio}

The sample was mounted on a continuous-flow He cryostat and photodoped with
a 100 nm diameter probe of a room temperature NSOM. The tip-to-sample
separation was controlled in the 5-10 nm range by means of a quartz
mechanical oscillator feed back system. Light from either a 10 mW, 
$\lambda $ = 632.8 nm HeNe laser or a 1 mW, $\lambda $ = 1.55 ${\rm \mu }$m
InGaAsP laser diode was coupled into the NSOM probe by means of a 2x2
optical fiber coupler. During
photogeneration the sample was illuminated only by the HeNe laser.
Reflectance changes were measured by illuminating the sample through the
NSOM probe (with only the 1.55 ${\rm \mu }$m laser light coupled into it)
and collecting the reflected light with a conventional microscope objective.\cite{AlGaAs} 
Photon fluxes per unit time in the near field were estimated to be 
$Q(632.8{\rm nm})\sim 3.5\times 10^{21}%
\frac{{\rm photons}}{\rm cm^2sec}$ and $Q(1.55{\rm \mu }{\rm m})\sim
2.5\times 10^{21}\frac{{\rm photons}}{\rm cm^2sec}$. For resistance
measurements the NSOM head was detached, the cryostat closed and pumped to 10$^{-6}$ Torr 
and the sample cooled to 120 K. It took less than 20 min  
between the end of photogeneration at room temperature and cooling the
sample to 120 K.

The main result of this paper is presented in Fig. \ref{f2}. The
NSOM reflectance scan, obtained after illuminating on a 1 $\mu$m 
side square for $t=2000$~sec, demonstrates $w \sim 
250$ nm confinement of
photogenerated carriers. The scan took 10 min to acquire, and was
measured from bottom to top in Fig. \ref{f2}. The different contrast between the
bottom and top sides of the square is due to the ongoing photocarrier
recombination. 

The time dependence of photocarrier recombination was further investigated
at the point marked by an arrow in Fig. \ref
{f2}. Raster scans accross the nanowire were averaged 100 times during a time span of 10 
sec
and the process was repeated every 30 sec. The results, shown in 
Fig. \ref{f3}a, \cite{irquench} are consistent with
the current understanding of photocarrier generation in RBCO. A trapped
electron has a probability $P_{i}\propto exp(-\Delta_s /k_{B}T)$ to become
ionized where $\Delta_s \sim 1$ eV.\cite{osqui1,julio} 
Room temperature recombination of electrons with holes from
the CuO planes decreases the available number of free carriers and,
hence, the optical conductivity at 1.55 ${\rm \mu }$m. Fig. \ref
{f3}b shows that the integrated intensity decreases at a rate slower than
exponential, as has been observed in wide area PPC experiments.\cite
{osqui1,julio} Although the dynamical range is too 
small to draw strong conclusions, the fact that the
decay is not a pure exponential suggests that even an area of (250 nm)$^{2}$
presents a distribution of $\Delta_s $. The most
important information about the time evolution is contained in Fig. \ref{f3}c, 
which shows the time dependence of the full-width half-maximum (FWHM) of
the line profile intensity. The FWHM is larger than the diameter of the NSOM
probe and increases only slightly more than the experimental error. As discussed
later, the initial width is associated with the diffusion of photocarriers 
before electron capture, and the small observed increase in width could be
associated with a much slower diffusion of trapped electrons.

To further investigate the properties of photogenerated carriers we have
studied the low temperature resistance of photogenerated wires. A 20 ${\rm \mu}$m 
long wire was written with the NSOM probe for 10 hours. After cooling the system to 
100 K the rest of the sample was photoexcited 
by means of a focused HeNe laser beam.
This last process was continued until no further change was observed in the DC
resistance. The low temperature resistance of the photoexcited pattern is shown in 
Fig. \ref{f4}a. The
drop in the curve, at T $\sim $ 29 K, corresponds to the superconducting
transition of the region photoexcited by the focused HeNe laser. At T $\sim $
10 K an additional three orders of magnitude drop is observed in the
resistance associated with the superconducting transition in the
NSOM-induced wire. A V-I characteristic measured at 4 K confirms this
result, as shown in Fig. \ref{f4}b. For comparison, we also show the V-I
characteristics of the photogenerated wire measured at 20 K as well as the
result expected for a copper wire of similar dimensions at room temperature.
Considering that our voltage noise was 70 nV, dissipation in
the sample at 4 K and I $\sim$ 9 ${\rm \mu }$A is more than three orders of
magnitude smaller than in copper, and more than four orders of magnitude
smaller than the sample at 20 K. The critical current, defined as the value where
the voltage is larger than the noise, gives I$_{c}\sim $ 9 ${\rm \mu }$A.
This gives a critical current density J$_{c}\sim 
10^{4}$ A/cm$^{2}$, comparable with J$_{c}$ for oxygen
depleted samples with similar T$_{c}$.

The most puzzling feature of the data arises from the difference
between the size of the NSOM tip and the width of the superconducting 
wire.\cite{experiment} 
Three main diffusive processes may be responsible for the
extra width: Diffusion of the {\it e-h} pairs before electron trapping, diffusion
of the trapped electrons, or diffusion of the conducting holes in the CuO 
planes. The extra holes pumped into the CuO planes move in the
potential well defined by the trapped electrons, and they can not fluctuate
more than a screening length, $l_{s}$. The density of photogenerated
carriers was estimated to be $n_{e}\sim 10^{20}\frac{1}{{\rm cm}^{3}}$\cite
{hall} leading to a Thomas-Fermi screening length $l_{s}\sim $ 1 nm.

The results of Fig. \ref{f3}c show that the other two processes are both 
present. Although {\it e-h} diffusion before electron capture appears to
account for most of the difference between probe and wire size, trapped
electrons remain marginally mobile, as evidenced by the upward curvature 
observed in Fig. \ref{f3}c. Thus localized electrons can hop randomly 
between trapping centers, similarly to the motion of carriers in
semiconductor impurity bands. The hopping rate will then
be given by

\begin{equation}
\tau^{-1} = \nu_{ph} exp(-\frac{\Delta_{s}}{k_BT}),  \label{eq1}
\end{equation}

\noindent where $\nu _{ph} \sim 10^{13}$ Hz is a typical phonon frequency.
The two dimensional diffusion distance after 1000 sec is

\begin{equation}
\langle d_{t}^{2}\rangle =(30nm)^{2}=4Nl^{2}\sim 4{\frac{1000{\rm sec}}{\tau }%
}d_{o}^{2},  \label{eq2}
\end{equation}

\noindent
where $N$ is the number of hops and $l$ is the hopping distance, which is
taken as the separation between trapping centers $d_{o}\sim 1$ nm, and 30 nm 
is the observed increase of the width at room temperature.

From Eq. (\ref{eq1}) and (\ref{eq2}), we get $\Delta _{s}\sim 850$ meV. This
value is comparable to the excitation energy measured by PPC decay
experiments\cite{osqui1,julio} suggesting a common source for both
processes. Since the captured electrons become thermally detrapped the {\it e-h} 
pairs can recombine. The recombination time was estimated to be 
$\tau _{R}\sim 10^{-9}$sec, \cite{Heeger} 
so that a trapping time $\tau _{t}\sim 4\times 10^{-12}$sec is implied.

For technological applications a better understanding of the diffusion
processes may allow control of the width of photoinduced wires. We believe that {\it e-h} 
plasma diffusion is the dominant mechanism for
determining the initial width of the wire, $d_{i}\sim $~200 nm. Considering
both hot electron diffusion and thermalized diffusion, we can express

\begin{equation}
\langle d_i^2 \rangle = 4 ( v_{th} l_{th} \tau_t + v_B l_h \tau_h),
\label{eq3}
\end{equation}

\noindent where $v_{B}$($v_{th}$) is the velocity of the hot (thermal) electrons, 
$l_{h}$($l_{th}$) their mean free paths, and $\tau _{h}$ is the
time required for the electronic system to thermalize. The observed width
cannot be explained with the first term alone since the low
thermal velocities of electrons ($v_{th}\sim 10^{6}~\frac{\rm cm}{\rm sec}$) would 
require
unreasonably large values for $l_{th}$. Band velocities are in the 
$10^{8}\frac{{\rm cm}}{{\rm sec}}$ range, however. For the hot carrier process to play
a significant role $\tau _{h}\sim 10^{-12}$sec is required. Although the thermalization
time has been estimated to be $10^{-13}$ sec,\cite{times} the agreement is not 
unreasonable in view of the uncertainties involved in the estimates.
However, these considerations suggest that diffusion may occur in a more
complicated fashion than represented by Eq. \ref{eq3}. Further studies of
the processes involved are under way and will be
published elsewhere.

In summary, a high T$_{c}$ nanowire has been realized for the first time
using NSOM photodoping. Photoholes transferred to the CuO planes
move in a potential well defined by trapped electrons. The high spatial
resolution of the NSOM probe provides new information about the physical
phenomena and the characteristic parameters involved in the capture of
photogenerated electrons. The nanowires, which width is determined by hot 
carrier diffusion, present 
enhanced optical conductivity and superconducting
properties similar to those observed in oxygen doped materials. NSOM writing
of high T$_c$ superconducting wires opens new avenues towards the fabrication of
superconducting devices.

The authors would like to thank F. Wellstood for useful discussions. Work partially 
supported by grants from
CONICET, Fundaci\'on Antorchas, and Fundaci\'on Balseiro. J. G. and E. O. 
are members of CONICET, Argentina.

\begin{figure}[tbp]
\caption{Reflectance of the photogenerated wire measured with the NSOM
at 1.55 $\mu$m. The scanning area is 2 $\times$ 2 $\mu$m$^2$. Maximum contrast is about 
10 \%.}
\label{f2}
\end{figure}

\begin{figure}[tbp]
\caption{Time evolution of the reflectance measured for raster scans across the wire. 
(a) Measured signal at 30
sec intervals. (b) Integrated intensity for each curve of (a). The dotted line 
represents a pure exponential behavior. (c) FWHM
of the data of (a).}
\label{f3}
\end{figure}

\begin{figure}[tbp]
\caption{Transport measurements of a photoinduced wire. a.- D.C. resistance
as a function of temperature. b.- V-I curves measured at ($\bullet$)  4 K and 
(---) 20 K on the
sample. ($\cdots$)Dissipation expected for a Cu wire of similar dimensions at room
temperature.}
\label{f4}
\end{figure}


\begin{references}
\bibitem{aplica}  See {\it Applied Superconductivity},
edited by H. C. Freyhardt (DGM Informationsgesellschaft, Oberursel, 1993),
and references therein.
\bibitem{PPC}  A. I. Kirilyuk, N. M. Kreines, and V. I. Kudinov, Pis'ma Zh. \'{E}ksp. 
Teor. Fiz. {\bf 52}, 696 (1990) [JETP Lett. {\bf 52}, 49 (1990)].
\bibitem{osqui1}  E. Osquiguil {\it et al.},
Phys. Rev. B {\bf 49}, 3675 (1994);  V. I. Kudinov {\it et al.}, Phys. Rev B {\bf 47},
9017 (1993).
\bibitem{gladys}  G. Nieva {\it et al.}, Appl. Phys. Lett. {\bf 60}, 2159
(1992).
\bibitem{schuller}  T. Endo {\it et al.},
Phys. Rev. B {\bf 54}, 3750 (1996).
\bibitem{hall}  G. Nieva {\it et al.}, Phys. Rev. B {\bf 46},
14249 (1992).
\bibitem{julio}  J. Guimpel {\em et al.}, Phys. Rev. B {\bf 56}, 3552 (1997)
\bibitem{NSOM}  See for example {\it Optics at the Nanometer Scale}, edited by M. 
Nieto-Vesperinas and N. Garc\'{\i}a, NATO ASI Series E (Kluwer, Dordrecht, 1996), 
and references therein.
\bibitem{AlGaAs}  R. S. Decca, H. D. Drew, and K. L. Empson, Appl. Phys.
Lett. {\bf 70}, 1932 (1997).
\bibitem{irquench}  D. C. Chew {\it et al.}, Appl. Phys. Lett. 
{\bf 69}, 3260 (1996), observed {\it IR} radiation quenching of
photogeneration. Reflectivity measurements at 1.55 $\mu $m laser could
lead to a decrease in the observed relaxation times.
\bibitem{experiment}  Previous experiments (Ref. \cite{AlGaAs}) indicate that lateral tip 
drift over several thousand seconds is less than 5 nm. 
\bibitem{Heeger}  G. Yu {\it et al.}, Phys. Rev. B {\bf %
45}, 4964 (1992).
\bibitem{times}  N. Bluzer, Phys. Rev. B {\bf 44}, 10222 (1991).
\end{references}
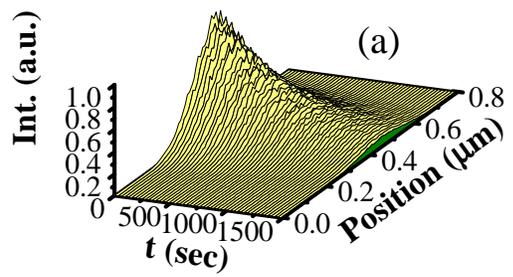
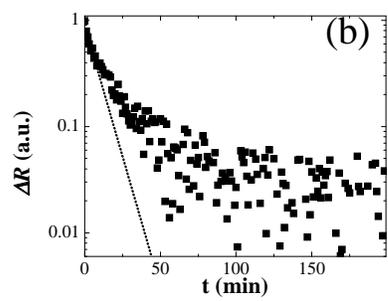
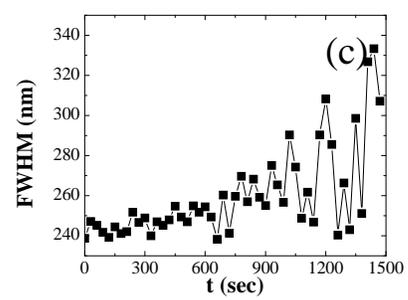
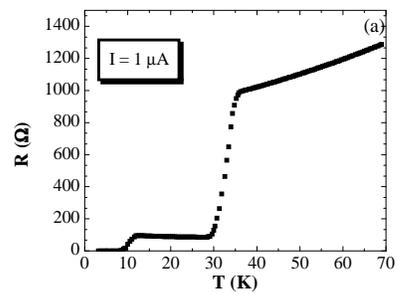
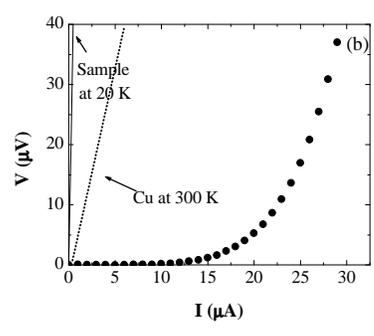
\end{document}